\documentstyle[amssymb,multicol,aps,prb,epsfig]{revtex}

\begin{document}
\title{{Calculation of magnetic anisotropy energy in SmCo$_{5}$ }}
\author{P. Larson, I.I. Mazin and D.A. Papaconstantopoulos}
\address{Center for Computational Materials Science\\
Naval Research Laboratory,
4555 Overlook Ave SW
Washington, DC 20375-5320
}
\date{\today}
\maketitle

\begin{abstract}
SmCo$_{5}$ is an important hard magnetic material, due to its large magnetic
anisotropy energy (MAE). We have studied the magnetic properties of 
SmCo$_{5}$ using density functional theory (DFT) calculations where the Sm 
$f$-bands, which are difficult to include in DFT calculations, have been
treated within the LDA+U formalism. The large MAE comes mostly from the Sm 
$f$-shell anisotropy, stemming from an interplay between the crystal field 
and the spin-orbit coupling.  We found that both are of similar strengths, 
unlike some other Sm compounds, leading to a partial quenching of the orbital 
moment ($f$-states cannot be described as either pure lattice harmonics 
or pure complex harmonics), an optimal situation for enhanced MAE. A smaller 
portion of the MAE can be associated with the Co-d band anisotropy, related 
to the peak in the density of states at the Fermi energy. Our result for the 
MAE of SmCo$_{5}$, 21.6 meV/f.u., agrees reasonably with the experimental 
value of 13-16 meV/f.u., and the calculated magnetic moment (including the 
orbital component) of 9.4$\mu_{B}$ agrees with the experimental value of 
8.9$\mu_{B}$.
\end{abstract}

\begin{multicols}{2}
The permanent magnet intermetallic compound SmCo$_{5}$ has been studied
extensively experimentally\cite{kirchmayr,radwanski} and theoretically\cite%
{malik,nordstrom,trygg,yamaguchi1,daalderop,yamaguchi2,steinbeck,steinbeck2,larson}. The interest in these materials is fueled by their large
magnetic anisotropy energy (MAE), which is defined as the difference between
the ground-state energies due to rotation of the magnetic field
(magnetization direction). It is generally understood that the main source
of the large MAE in SmCo$_{5}$ and other Sm-Co magnets is large magnetic 
single-site anisotropy of the Sm 
$f$-shell\cite{steinbeck,richter,novak,hummler}. In simple terms, this means  
the strong spin-orbit coupling tries to align the Sm $f$-shell with the 
magnetic field, causing the $f$-shell to rotate with the field. Sm atoms in a 
lattice interact with the crystal field, and the energy of this interaction 
depends on the orientation of the Sm $f$-shell in the lattice. This is the 
leading contribution to the MAE. Note that if the crystal field is too small, 
the $f$-shell rotates freely with the magnetic field, producing no MAE, while
if the spin-orbit is too small, the orbital moment is quenched and again
no MAE appears. We will see below that in SmCo$_{5}$ both interactions
are comparable, producing a large MAE.

This basic understanding has existed for long time, and has been the basis
of several model calculations, where atomic calculations for Sm have been 
combined with the crystal field parameters derived from first principles 
calculations\cite{fahnle}. However, first principles calculations which could
provide a quantitative analysis of different components of MAE in SmCo$_{5}$
are still missing, to the best of our knowledge. In this paper we report
such calculations, using an all-electron, full-potential, relativistic
linearized augmented plane wave (FLAPW) method with an LDA+U extension to 
account for Coulomb correlations in the $f$-shell.

Our analysis is organized as follows: We start by looking at the ``Co part'' 
of the MAE, the part {\it not} related to the Sm single-site anisotropy. We 
do this by investigating YCo$_{5}$ which forms in the same crystal structure 
as SmCo$_{5}$ but contains no $f$-electrons. We calculate the MAE for 
YCo$_{5}$ and for the hypothetical Co$_{5}$ compound (corresponding to 
YCo$_{5}$ with Y removed) to elucidate the role of Y, and analyze the 
electronic origin of the MAE using the force theorem. We then move to 
SmCo$_{5}$, where we first compute the MAE as a function of the Coulomb 
repulsion parameter U. These results demonstrate the nature of the Sm 
$f$-states which shows strong competition between the crystal field and the 
spin-orbit interactions.

For our electronic structure calculations we have used the self consistent 
FLAPW method\cite{singh}. The Local Density Approximation (LDA) of Perdew and 
Wang\cite{perdewwang} and the Generalized Gradient Approximation (GGA) of 
Perdew, Burke, and Ernzerhof\cite{perdew} were used for the correlation and 
exchange potentials. Calculations were performed using the WIEN 
package\cite{wien2k}. Local orbital extensions\cite{LAPW1} were included in 
order to accurately treat the upper core states and to relax any residual 
linearization errors. A well converged basis consisting of approximately 300 
LAPW basis functions in addition to the local orbitals was used with the Y 
and Sm sphere radii set to 2.115 a.u. and the Co sphere radii to 2.015 a.u.. 
The results varied only within a few percent for reasonable choices of atomic 
radii (2.0-3.0 a.u.). In our previous study\cite{larson} on YCo$_5$ we 
established that the plane-wave cut-off parameters $RK_{MAX}$ and $G_{MAX}$ 
of 9 and 14, respectively, are suffucient for MAE calculations,so the same 
parameters were used in this work.Spin-orbit (SO) interaction was 
incorporated using a second variational procedure\cite{spinorb}, where all 
states below the cutoff energy 1.5 Ry were included. The most recent version 
of the WIEN code, WIEN2k, includes the so-called $p_{1/2}$ 
extension\cite{singh,kunes}, which accounts for the finite character of the 
wave function at the nucleus for the $p_{1/2}$ state that cannot be 
adequately represented as a linear combination of a finite number of 
solutions of the radial Schr\"{o}dinger equation with $l=1$.

\begin{figure}[htb]
\centering
\epsfig{file={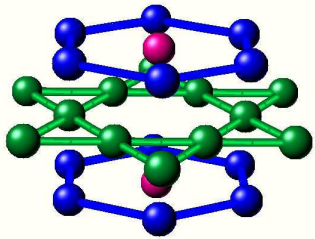}, width=8 cm}
\caption{Crystal structure of SmCo$_{5}$. Shown are 2 layers of hexagonal 
Co(2c) with a Kagome lattice of Co(3g) atoms lying between. The Sm atoms lie 
in the middle of the hexagons in the Co(2c) layers.}
\label{cryst}
\end{figure}

The crystal structure of SmCo$_{5}$ and YCo$_{5}$ is that of CaCu$_{5}$ 
($P6/mmm$, No. 191). The experimental values of $a$ and $c/a$ used in the 
calculation are 9.452 a.u. and 0.792 for SmCo$_{5}$ and 9.313 a.u. and 0.806 
for YCo$_{5}$. The Co sites are separated into two sets of inequivalent 
atoms, Co(2c) having 2-fold multiplicity and Co(3g) having 3-fold 
multiplicity. (Figure \ref{cryst}) Including spin-orbit coupling into the 
calculation lowers the symmetry when the field lies along the plane (to 
$Pmmm$, No. 47), separating the 3 atoms corresponding to Co(3g) into two 
inequivalent sites which have multiplicities of 2 and 1, respectively. To 
eliminate a systematic error\cite{larson} we performed the calculation for 
both magnetization directions using the same, highest {\it common} symmetry 
group ($Pmmm$, in this case).\cite{wyckoff}

In our previous work\cite{larson}, we found the MAE of YCo$_{5}$ to be 0.32 
meV/Co, which is 0.44 meV/Co lower than the experimental number of 0.76 
meV/Co.\cite{buschow}. What is important for our discussion is that both 
experimental and calculated numbers are substantially larger than the MAE of 
the hcp Co (0.065 meV/Co)\cite{paige}. The question arises whether this 
enhancement is due to a different mutual arrangement of the Co ions, a charge 
transfer between Y and Co, or the MAE associated with the Y ions. To answer 
this question, we performed calculations for a hypothetical Co$_{5}$ compound 
defined as YCo$_{5}$ with Y removed with the positions of all cobalt atoms 
unchanged. We found MAE of 0.28 meV/Co (compared to 0.32 meV/Co in YCo$_{5}$), 
proving that the reason for the relatively large MAE in YCo$_{5}$ is a 
favorable arrangement of the Co atoms.

\begin{figure}[htb]
\centering
\epsfig{file={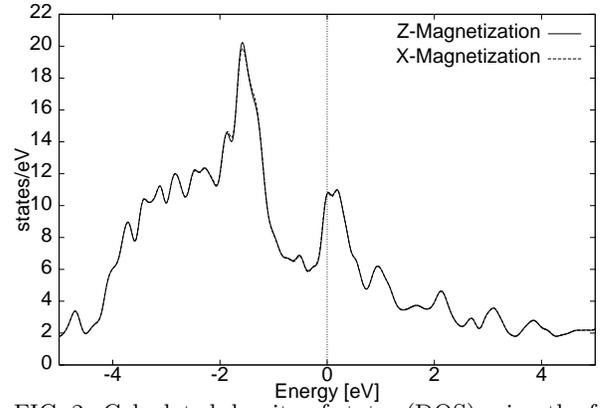}, width=8 cm}
\caption{Calculated density of states (DOS) using the force theorem for 
magnetization along the Z-axis and along the X-axis in YCo$_{5}$. The 
difference is only noticeable around the peak near -1.5 eV.}
\label{force}
\end{figure}

\begin{figure}[htb]
\centering
\epsfig{file={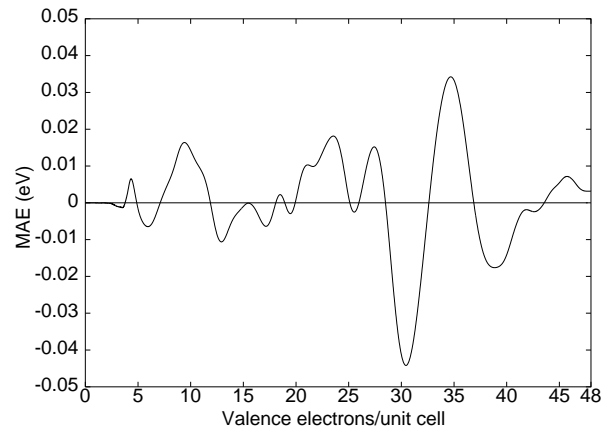}, width=8 cm}
\caption{Using the calculated density of states (DOS) of YCo$_{5}$ (Figure 2) 
for the two magnetization directions, the product of the energy and the 
difference in the DOS are plotted as a function of valence electrons/unit 
cell. At full occupation, 48 valence electrons, a small ($\sim$ 1.5 meV/f.u.) 
positive contribution to the MAE is related to the Co d-peak near the Fermi 
energy.}
\label{mae}
\end{figure}

We can actually pinpoint the microscopic origin of this large MAE. The MAE 
depends on subtle differences in the electronic structure under rotations of 
the external magnetic field. While our actual calculations used two 
self-consistent energies for the two field directions, it is convenient to 
analyze the results using the force theorem\cite{weinert}. According to this 
theorem, one can start from the same charge density (which is converged 
without spin-orbit), and then apply the spin-orbit within one iteration 
for the two different magnetization directions, {\bf e}$_{1}$ and 
{\bf e}$_{2}$. One then sums the eigenvalues for the occupied states in 
both cases, the difference of these sums corresponding to the MAE, good to 
second order in the change in the electron charge density.
\begin{eqnarray}
MAE \approx \Sigma_{i}^{occ} \epsilon_{i}({\bf e}_{1}) - 
\Sigma_{i}^{occ} \epsilon_{i}({\bf e}_{2}). \label{1}
\end{eqnarray}%
We used the force theorem to produce the DOS plots for YCo$_{5}$ shown in 
Figure \ref{force}. The two plots are practically indistinguishable on 
this scale, except near -1.5 eV, which shows a slight variation near the top 
of the peak. Also notice the smaller Co $d$-band minority peak which crosses 
the Fermi level on its left side. We now plot the corresponding difference in 
the one-electron energy (Eq. \ref{1}) against the number of valence electrons
(Figure \ref{mae}). Below full occupation (48 electrons), there are positive 
and negative variations to the running value of the MAE which generally 
average to zero. At full occupation, 48 valence electrons, the MAE shows a 
small positive contribution ($\sim$ 1.5 meV/f.u.), in agreement with the 
results found by taking the energy difference of two self-consistent 
calculations. This corresponds to the differences under rotation of the 
magnetization direction in the Co $d$-band minority peak which is cut by the 
Fermi level (Figure \ref{force}). Dopant atoms change the DOS and, hence, the 
calculated MAE enough so that one cannot blindly trust the results of a rigid 
shift of the Fermi energy, but this plot indicates that small amounts of Fe 
doping may initially increase the MAE before it decreases and changes sign, 
as has been seen experimentally\cite{rothwarf} and theoretically\cite{unpub}.

Several difficulties are encountered going from YCo$_{5}$ to SmCo$_{5}$. 
Computationally, the main difference between the two compounds is that Sm 
includes an open $f$-shell that cannot be described in the framework of the 
conventional LDA theory. Indeed, uncorrected DFT (whether LDA or GGA) 
calculations incorrectly pin all of the $f$-orbitals at the Fermi energy 
($E_{F}$) in SmCo$_{5}$ and other systems containing unfilled $f$-orbitals. In 
order to circumvent this problem, the previous electronic structure 
calculations\cite{steinbeck,richter,novak,hummler} for SmCo$_{5}$ did not 
treat the $f$-orbitals as valence electrons but as unhybridized core 
electrons. However, the width of the $f$-bands is $\sim$ 1 eV, much too 
large for this approximation. Since the single-site MAE of Sm is due to the 
asphericity of the $f$-shells\cite{buck}, in other words, to the interaction 
of this shell with the crystal field, understanding the exact shape and 
occupation of the $f$-orbitals is crucial. Therefore we applied a Hubbard $U$ 
correction to the $f$-orbitals which naturally splits the $f$-bands into 
lower and upper Hubbard bands with nearly-integer occupancies.

While the LDA+U method seems the proper procedure for handling the localized 
Sm $f$-orbitals, the question remains which form of the LDA+U should be used. 
There are several prescriptions, differing mainly in the way the double 
counted energy components are subtracted off\cite{MP}. The two most common 
are referred to in the WIEN code\cite{wien2k} as SIC and AMF (the former 
name is misleading from the physical point of view, so Mazin 
{\it et al}\cite{MP} suggested an acronym FLL, fully localized limit). As 
discussed by Mazin {\it et al}\cite{MP}, the FLL prescription is more 
appropriate for well localized orbitals, such as $f$-electrons in SmCo$_{5}$, 
so it was used in our calculations\cite{note1}.

The LDA+U method includes two parameters, $U$ and $J$, that have no rigorous 
definition in a solid. In an atom, one way to define $U$ and $J$ is {\it via} 
the derivatives of the energy of the atomic level with respect to their 
occupancies\cite{olle}:%
\begin{eqnarray}
U_{f} &=&\frac{\partial \varepsilon _{f\uparrow }}{\partial n_{f\downarrow }} \\
J_{f} &=&\frac{\partial (\varepsilon _{f\uparrow }-\varepsilon _{f\downarrow
})}{\partial (n_{f\uparrow }-n_{f\downarrow })}.
\end{eqnarray}%
We have applied these formulas to a quasiatom residing in a potential well 
defined by the DFT crystal potential, which is especially easy within the LMTO 
technique (the details of this procedure will be reported elsewhere). This 
gives for SmCo$_{5}$ $U_{f} \approx$ 5.2 eV and $J_{f} \approx$ 0.75 eV, 
which, as the subscripts explain, will act on the Sm $f$-orbitals in the DFT 
calculation. Note that this approach overscreens the Coulomb interaction by 
forcing all screening charge into one atomic sphere, therefore such 
calculated $U_{f}$ may be slightly underestimated.

\begin{figure}[htb]
\centering
\epsfig{file={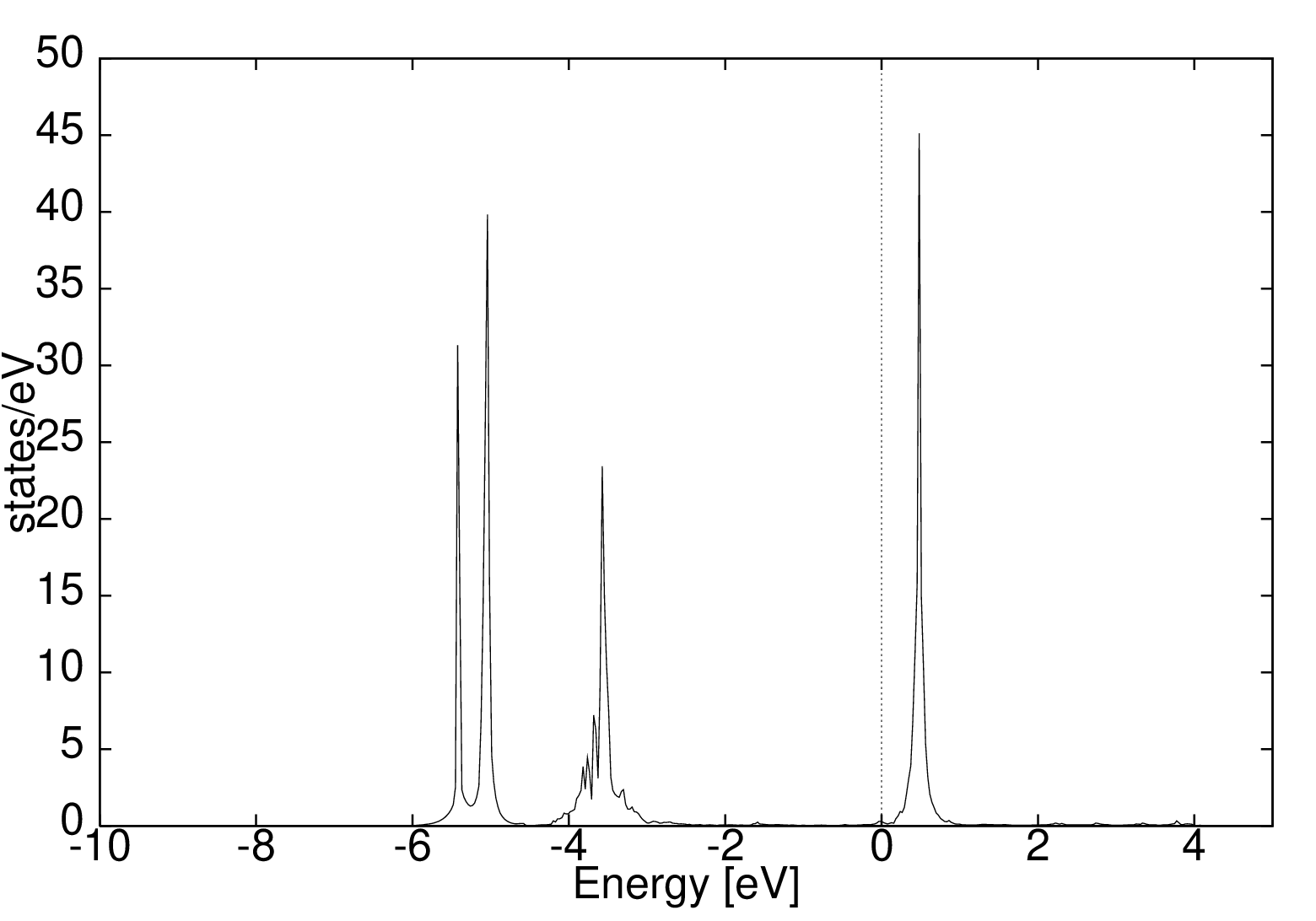}, width=8 cm}
\epsfig{file={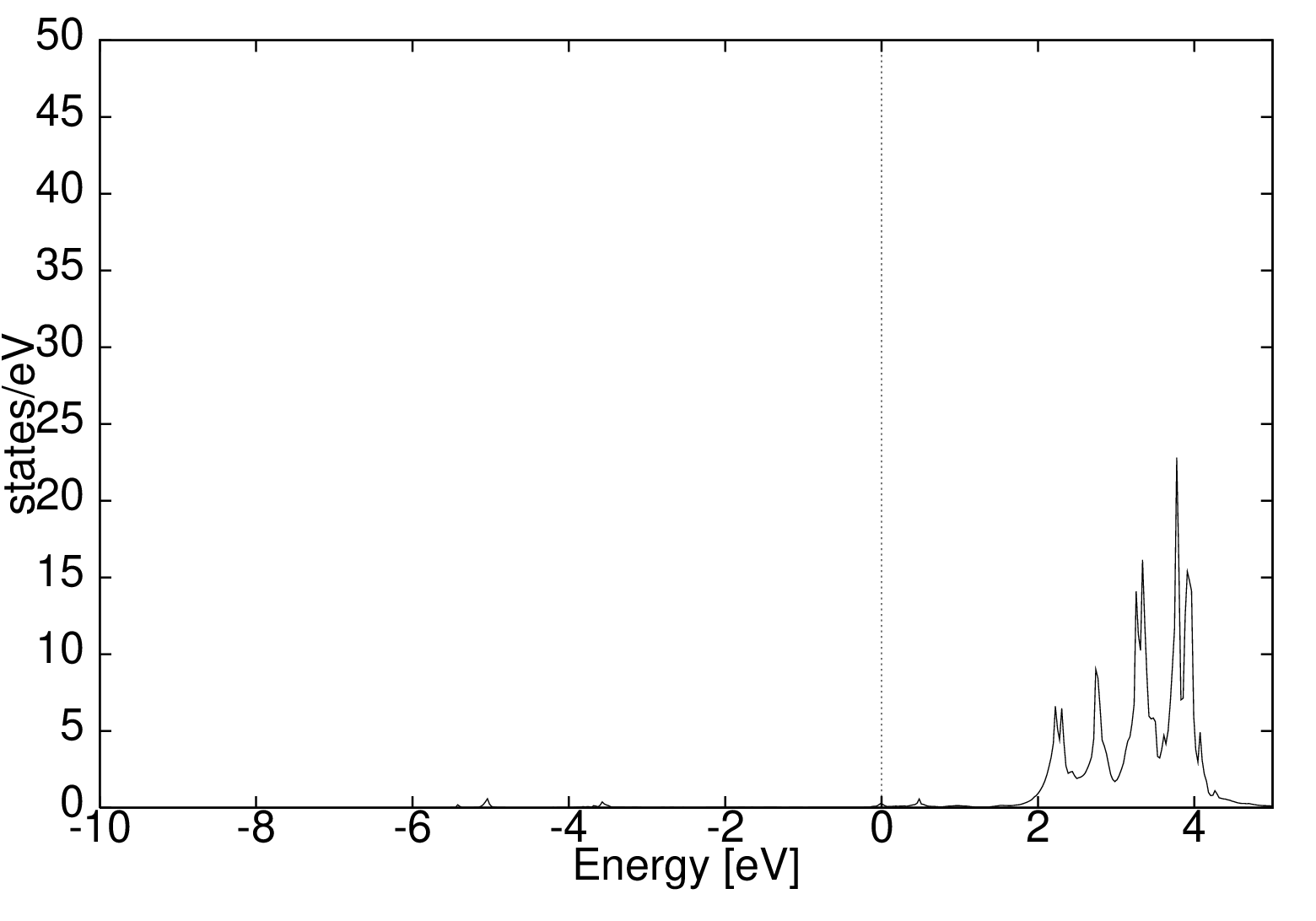}, width=8 cm}
\caption{Calculated (a) spin up and (b) spin down density of states (DOS)
for the Sm $f$-orbitals with GGA and LDA+U (SIC-LDA, GGA) in SmCo$_{5}$ 
including spin-orbit. The lowest spin-up peak corresponds to m = +3, the 
second to a combination of m = +2, +1.  The highest valence band peak is 
composed of m = 0, +2, -2 and m = +1, -1, +3, -3. The conduction band peak is 
composed of m = -2 and m = -3, -1.}
*\label{dosso}
\end{figure}

The effect of including LDA+U in the calculation can be seen by looking at 
the spin-up (Figure \ref{dosso}a) and spin-down (Figure \ref{dosso}b) 
contribution of the Sm $f$-orbitals to the density of states (DOS). With just 
DFT-GGA, the Sm $f$-orbitals form a narrow band pinned at $E_{F}$ (not 
shown). The addition of $U$ and $J$ shifts the unoccupied spin-minority 
$f$-band up by approximately ($U$ - $J$)/2 and splits the spin minority 
$f$-band into the lower and the upper Hubbard bands, separated roughly by $U$ 
- $J$. Furthermore, LDA+U enhances the Hund's second rule coupling\cite{hund2} 
and tries to unquench the orbital moment by making the orbital moment 
projection $m$ a good quantum number. Without spin orbit there is no Hund's 
second rule coupling, and orbital moment is quenched, that is, each $f$-band 
includes the same amount of $m$ = 3 character as of the $m$ = -3 character 
etc. Spin orbit in combination with Hubbard $U$ favors pure $m$-states. This 
scenario holds, for instance for such Sm compounds as SmAl$_{2}$ and SmZn, 
where the individual spin and orbital moments are known 
experimentally\cite{smalexp}, and the orbital moment is between -4.5 and -5 
$\mu _{B}$, roughly as expected from Hund's second rule, in agreement with 
LDA+U calculation with a sufficiently large $U$.\cite{smalthe}. In 
SmCo$_{5}$, however, the crystal field is sufficiently strong to prevent 
total unquenching, and as a result the $f$-occupation matrix is not diagonal 
in the $m$ representation (Table I), and the orbital moment is reduced to 
-2.8 $\mu _{B}$ (within the calculation).

\begin{figure}[htb]
\centering
\epsfig{file={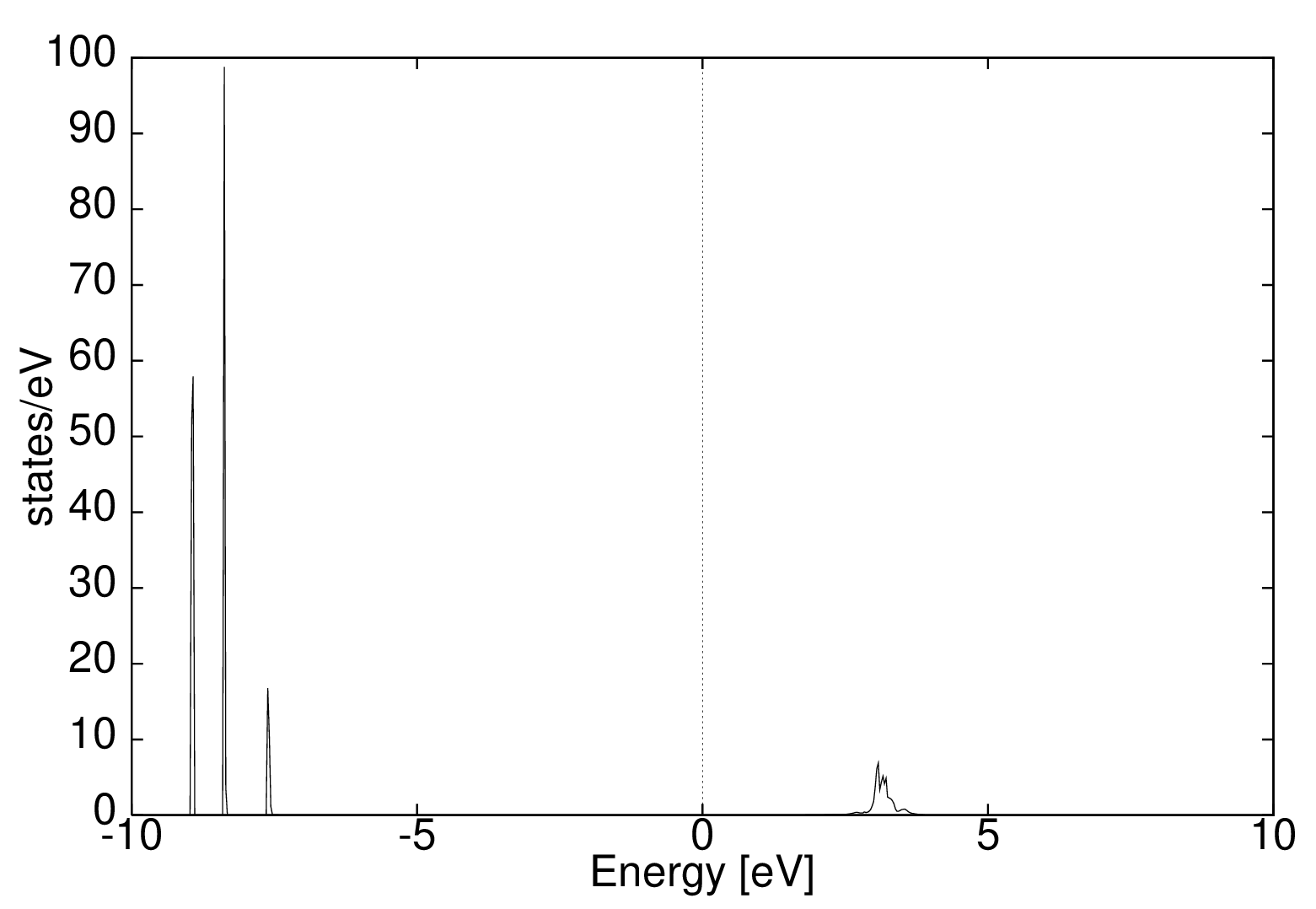}, width=8 cm}
\epsfig{file={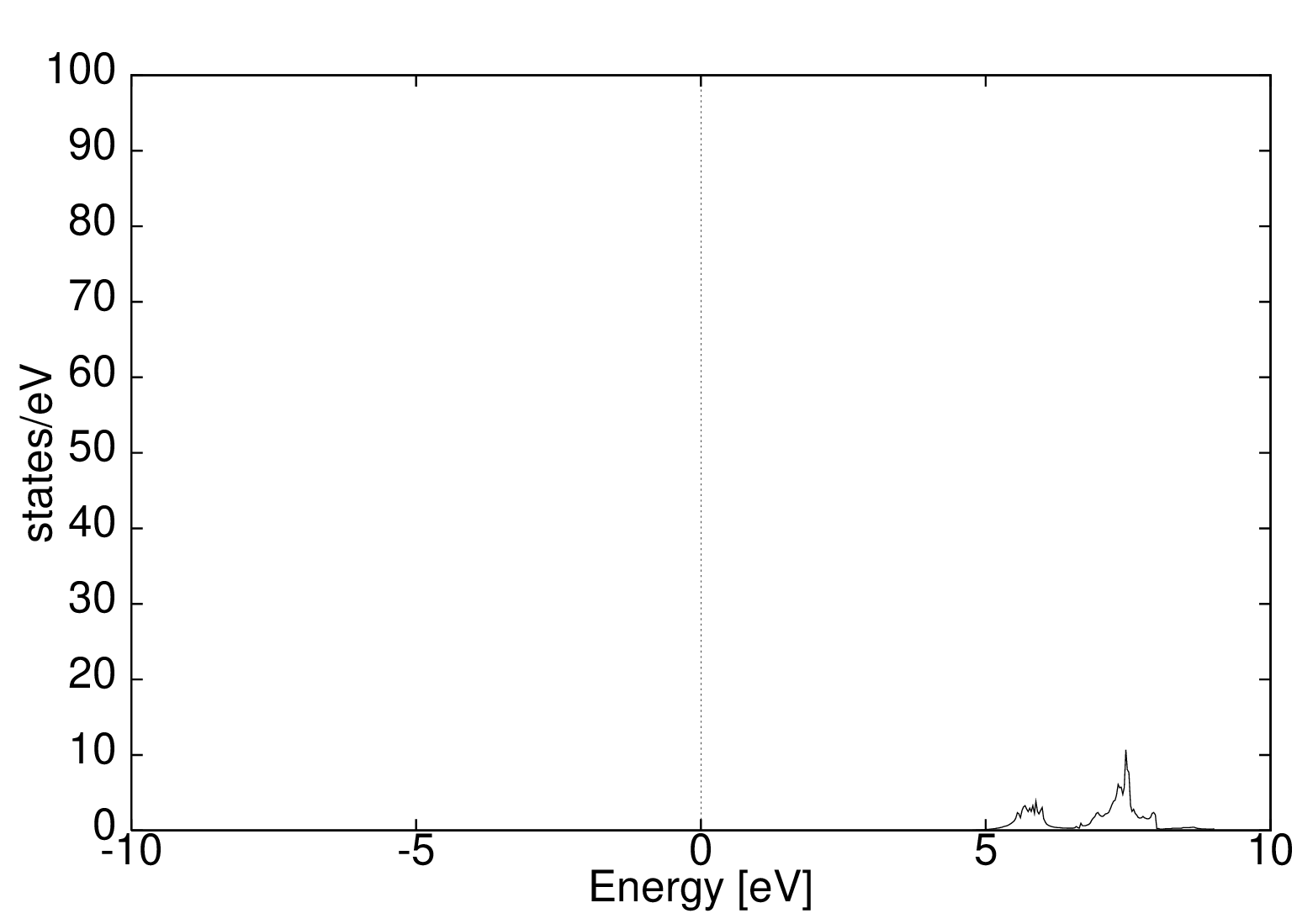}, width=8 cm}
\caption{Calculated (a) spin up and (b) spin down density of states (DOS)
for the Sm $f$-orbitals with GGA and LDA+U (SIC-LDA, GGA) in SmCo$_{5}$
{\it without} spin-orbit. The occupied bands consist of, correspondingly, 
the $E_{1u}$, the $B_{1u}$ and $B_{2u}$, and the $A_{2u}$ staes while the 
conduction bands consist of the two $E_{2u}$ states.}
\label{dos}
\end{figure}

This can be also illustrated by the DOS for SmCo$_{5}$, and comparing with
the $f$-occupation matrix (Table I). The lowest peak in the DOS (Figure 
\ref{dosso}a) is clearly dominated by $m$ = 3, corresponding to the first 
state in the Table. The next peak includes two states: a pure $m$ = 2 state 
and a nearly pure $m$ = 1 state. The third peak is also composed of two 
closely lying peaks, one of character $m$ = 0, +2, -2, and a second of 
character $m$ = +1, -1, +3, -3. The conduction band peak also contains two 
closely lying peaks, one with mostly $m$ = 0 and an admixture of $m$ = -2, and 
the second of $m$ = -1 with an admixture of $m$ = -3. The two empty states are 
very close in energy and correspond, one to $m$ = -2 with an admixture of 
$m$ =0, and the other to $m$ = -3 with an admixture of $m$ = -1. Note that if 
we apply LDA+U to {\it scalar relativistic} calculations, the charge state of 
the $f$-shell remains the same, but now the bands are formed by the real 
lattice harmonics, with the two $E_{2u}$ states forming the unoccupied upper 
Hubbard bands, and, correspondingly, the $E_{1u}$, $B_{1u}$, $B_{2u}$, 
and $A_{2u}$ states forming the occupied bands. The distance between the 
lowest ($E_{1u})$ and the highest ($A_{2u})$ occupied states gives us a gauge 
of the crystal field strength: $\sim 0.1$ Ry. (Figure \ref{dos})

A technical problem with LDA+U calculations is that, unlike conventional DFT
calculation, localized orbitals like $f$-states can converge to a number of 
metastable configurations\cite{smalthe}. There is no guarantee that the 
configuration shown in Fig. \ref{dos}a is the true ground state. However, we 
used different starting configurations but were never able reach a 
self-consistent configuration with pure $m$-states. The calculated magnetic 
moment, which depends crucially on the orbital configuration, agrees with 
experiment. The calculated spin moment is 12.2 $\mu _{B}$, and the Sm orbital 
moment is -2.8 $\mu _{B}$, while the Co orbital moments are about 0.1 
$\mu _{B}$ each, leading to the total moment of 9.9 $\mu _{B}$, to be 
compared with the experimental value of 8.9 $\mu_{B}$\cite{zhao2,deportes}. 
As mentioned, fully unquenched orbital moment would be between -4.5 and -5 
$\mu _{B}$\cite{smalthe}, thus reducing the total moment to 7.2-7.7 
$\mu _{B}$ and increasing the disagreement with the experiment by a factor of 
2-3.

The calculated MAE of SmCo$_{5}$ comes from two sources. One is the MAE of 
the Co sublattice, analogous to that in YCo$_{5}$. The other is the 
single-site anisotropy of the Sm $f$-shell. The strong spin-orbit effect on 
the Sm $f$-shell is necessary due to the small MAE resulting when the 
spin-orbit effects are weak compared to crystal field effects, as in 
YCo$_{5}$. If there were no crystal field effects in SmCo$_{5}$, the 
$f$-states would be pure $m$-states, so the direction of the orbital 
quantization axis would always coincide with the magnetic field, by virtue of 
the spin-orbit interaction. Assuming a crystal field interaction much weaker 
than the spin-orbit, we observe that the energy of the $f$-shell with its 
orbital moment aligned along 001 or along 100 comes from the dependence of 
the crystal field energy on the orientation of the $f$-shell. The stronger 
the crystal field, the larger the MAE. The fact that the calculated $f$-bands 
in SmCo$_{5}$ are not pure $m$-states indicates that the crystal field in 
this compound is strong, comparable with the spin-orbit interaction. In fact, 
our preliminary calculations for a sister compound, 
Sm$_{2}$Co$_{17}$\cite{2-17} indicate that the $f$-states there are much 
closer to pure $m$-states than those in SmCo$_{5},$ that is, that the crystal 
field there is weaker, in agreement with the reduced MAE per Sm in 
Sm$_{2}$Co$_{17}$.\cite{217exp}

As discussed above, the single site MAE is defined by a delicate balance 
between the crystal field, the spin-orbit interaction, and the Hubbard 
repulsion. Since the LDA+U method implements the principal aspects of the 
latter, we do expect to have a reasonable description of MAE within this 
method. However, the accuracy of such calculations is necessarily limited. A 
good understanding of this can be gained by comparing the MAE calculated 
within two different versions of the WIEN code. Using the WIEN2k 
package\cite{wien2k}, in WIEN2k.01 the Hubbard correction is applied 
{\it before} solving the second-variational spin-orbit equations, while in 
the WIEN2k.02 version it is done {\it simultaneously}. We obtain an MAE of 
12.6 meV/f.u. using the former version, and 21.6 using the latter. Both 
compare favorably with the experimental number of 
13-16 meV/f.u.\cite{szpunar,zhao,pelecky,chen,omari}, emphasizing, however, 
the sensitivity of the result to the treatment of the correlation effects.

It is worth noting that {\it without} the LDA+U correction, in DFT-GGA, the 
calculated MAE was $-11.1$ meV/f.u. (note the wrong sign: the easy axis in 
the plane rather than axial), with the spin moment of 12.9 $\mu _{B}$ and an 
orbital moment of -1.5 $\mu _{B}$. Also, calculations from the Dresden 
group\cite{steinbeck}, otherwise similar to ours (we agree on the MAE in 
YCo$_{5}$\cite{larson}), but with $f$-electrons treated as an 
\textquotedblleft open core\textquotedblright, produced MAE of only 
$\sim$ 7meV/f.u.; presumably, such 
\textquotedblleft atomic\textquotedblright\ description of the $f$-electrons 
underestimate the crystal field effects.

Since the parameter $U$ is not very well defined, it is always instructive 
to check the dependence of the results on $U$, particularly since the 
above-mentioned quasiatomic procedure tends to underestimate $U$. Indeed, we 
found that if $U$ is reduced to $\approx$ 4 eV the upper Hubbard bands are 
too close to the Fermi level and the MAE is substantially overestimated 
($\sim$ 40 meV/f.u.). On the other hand, when $U$ was increased to 6.0 eV the 
MAE, as expected, changed little. It is also worth noting that using 
LDA\cite{perdewwang} instead of GGA\cite{perdew} improves the total magnetic 
moment (8.6 $\mu _{B}$, with the experimental value being 8.9 $\mu _{B}$), 
but worsens the MAE (26.0 meV/f.u., to be compared with 21.6 meV/f.u. in GGA 
for the same value of $U$, or with the experiment, 13-16 meV/f.u.). 
Interestingly, both in YCo$_{5}$ and SmCo$_{5}$ the GGA results are in better 
agreement with the experiment for the MAE than LDA, although in the former 
compound calculations underestimate MAE, and in the latter overestimate. 

To conclude, we have performed first principle calculations of the magnetic 
properties of SmCo$_{5}$ using a highly accurate LAPW code including the LDA+U 
formalism. We obtained much better agreement with experiment than previous 
methods which treated the $f$-orbitals as open core rather than valence 
states. Comparing the calculation for SmCo$_{5}$ with YCo$_{5}$ and with the 
hypothetical $\square $Co$_{5}$ we conclude that the MAE of the Co 
sublattice comes from a favorable arrangement of the Co atoms, which leads 
to a peak in the spin-majority DOS at the Fermi level. The MAE of the 
Sm $f$-shell comes from the interplay between the spin-orbit coupling, which 
tends to align the $f$-shell according to the magnetization direction and 
the crystal field, that aligns it according to the crystal lattice. In 
SmCo$_{5}$ both interactions appear to be of approximately the same 
strength, which makes SmCo$_{5}$ such an exceptionally hard magnet even 
compared with other Sm-Co compounds.

This work was supported by the Office of Naval Research and DARPA Grant No. 
63-8250-02.

\begin{table}[tbp]
\caption{$f$-occupation matrix for SmCo$_5$ ($U=$5 eV, magnetization 
direction is 001). The first column lists the eigenvalues, the other
the corresponding eigenvectors in terms of spherical harmonics with
give $m$.}
\label{tbl}
\begin{tabular}{c|ccccccc}
e-value & 3 & 2 & 1& 0 & $-1$& $-2$ &$-3$\\
0.96&0.81&0&0.05&0&0&0&0.14\\
0.95&0&1.&0&0&0&0&0\\
0.95&0.12&0&0.79&0&0&0&0.09\\
0.95&0.01&0&0.04&0&0.77&0&0.17\\
0.94&0&0&0&0.66&0&0.35&0\\
0.04&0.06&0&0.12&0&0.22&0&0.61\\
0.02&0&0&0&0.35&0&0.66&0\\
\tableline
\end{tabular}
\end{table}
\end{multicols}

\begin{references}
\bibitem{note1} As a test, we also performed calculations using LDA+U-AMF;
the results for the orbital moment were considerably worse (-1.5$\mu _{B}$
compared to -2.8$\mu _{B}$) and the calculated MAE was heavily overestimated.
\end{references}
\end{document}